\documentclass[twocolumn,
showpacs,preprintnumbers,amsmath,amssymb]{revtex4}

\usepackage{amssymb,amsmath,amsthm}
\usepackage{epsfig}
\usepackage{multirow}
\newtheorem{Thm}{Theorem}

\newtheorem{Prop}{Proposition}

\theoremstyle{definition}

\newcommand{\bra}[1]{{\left\langle #1 \right|}}
\newcommand{\ket}[1]{{\left| #1 \right\rangle}}

\newcommand{\B}{\mbox{$\mathbb B$}}
\newcommand{\X}{\mbox{$\mathbb X$}}

\newcommand{\T}{\mbox{$\mathrm{tr}$}}

\begin{document}
\title{Strong polygamy of multi-party $q$-expected quantum correlations}

\author{Jeong San Kim}
\email{freddie1@khu.ac.kr} \affiliation{
 Department of Applied Mathematics and Institute of Natural Sciences, Kyung Hee University, Yongin-si, Gyeonggi-do 446-701, Korea
}
\date{\today}

\begin{abstract}
We show that the polygamous nature of multi-party quantum correlations can be characterized in a {\em stronger} form based on Tsallis $q$-entropy and $q$-expectation value. By considering the amount of entanglement that can be distributed in multi-party systems, we establish a class of strong polygamy inequalities of multi-party entanglement in terms of Tsallis $q$-entropy and $q$-expectation for $q \geq 1$. Our new class of inequalities is in fact tighter than the usual polygamy inequalities of multi-party entanglement, and the tightness is explicitly illustrated by an example. Moreover, our new class of inequalities is concerned with the $q$-expected entanglement distributed between a single party and any possible subsets of the rest parties whereas the usual polygamy inequality only considers the entanglement between a single party and another. We further establish the equivalence between strong polygamy inequalities of quantum entanglement and quantum discord distributed in multi-party quantum systems.
\end{abstract}

\pacs{
03.67.Mn,  
03.65.Ud 
}
\maketitle
\section{Introduction}

One distinct property making quantum mechanics fundamentally different from classical one is
in the way of sharing correlations; whereas classical correlations can be shared among parties with
no certain restriction, quantum correlations have limited shareability and distribution in multi-party
quantum systems. For example, two parties $A$ and $B$ that are in maximally entangled state
cannot be entangled with any third party $C$. This restricted shareability of multi-party quantum entanglement is referred to as the {\em monogamy of entanglement}~\cite{KGS}.

Monogamy of entanglement was mathematically characterized in forms of an inequality; for a three-qubit pure state $\ket{\psi}_{ABC}$ with its two-qubit reduced density matrices $\rho_{AB}=\T_C \ket{\psi}_{ABC}\bra{\psi}$ and $\rho_{AC}=\T_B \ket{\psi}_{ABC}\bra{\psi}$, Coffman, Kundu and Wootters(CKW) established a monogamy inequality~\cite{ckw}
\begin{equation*}
\tau\left(\ket{\psi}_{A(BC)}\right)\geq \tau\left(\rho_{AB}\right)+\tau\left(\rho_{AC}\right),
\label{ckw}
\end{equation*}
where $\tau\left(\ket{\psi}_{A(BC)}\right)=4 \det \rho_A$ is the {\em tangle}(or square of concurrence) of $\ket{\psi}_{ABC}$ with respect to the bipartition between $A$ and $BC$, and $ \tau\left(\rho_{AB}\right)$ is the tangle of $\rho_{AB}$ defined by the minimum average tangle
\begin{equation*}
\tau\left(\rho_{AB}\right) = \min \sum_i p_i \tau(\ket{\psi_i}_{AB})
\label{tau}
\end{equation*}
over all possible pure-state decompositions of $\rho_{AB}=\sum_{i} p_i |\psi_i\rangle_{AB}\langle\psi_i|$.
Later, CKW inequality was generalized for multi-qubit systems as well as some classes of multi-qudit systems based on various entanglement measures~\cite{ov,kds}.

CKW inequality characterizes the mutually exclusive nature of bipartite entanglement shared in multi-party quantum systems. On the other hand, {\em assisted entanglement}, that is a dual quantity of entanglement measure, is known to have a dually monogamous property in multi-party quantum systems. 
Using {\em tangle of assistance}
\begin{equation*}
\tau^a\left(\rho_{AB}\right) = \max \sum_i p_i \tau(\ket{\psi_i}_{AB})
\label{tau}
\end{equation*}
that takes the maximum average tangle over all possible pure-state decompositions of $\rho_{AB}$,
three-qubit state $\ket{\psi}_{ABC}$ was shown to have the following dual monogamy inequality~\cite{GBS}
\begin{equation*}
\tau\left(\ket{\psi}_{A(BC)}\right)\leq \tau^a\left(\rho_{AB}\right)+\tau^a\left(\rho_{AC}\right).
\label{ckw}
\end{equation*}
It is clear that tangle of assistance is a dual to tangle because one takes the maximum while the other takes the minimum. Moreover, this dual quantity of tangle also provides a dual monogamy(thus {\em polygamy}) inequality of three-party entanglement. Later polygamy inequality was generalized into multi-party quantum systems~\cite{GBS,BGK,Kimpolys}.

If two parties $A$ and $B$ are maximal entangled, it is also known that they cannot be even classically correlated with $C$.
Conversely, maximal classical correlation between $A$ and $B$ also forbids them from sharing entanglement with $C$~\cite{KW}. Thus a perfect correlation(classical or quantum) between two systems can even exclude the possibility of sharing different kind of correlations with other systems. Other correlations such as Bell nonlocality~\cite{SG, CH} and quantum discord~\cite{BZYW} are
also known to have some trade-off relation in their shareability and distribution.

The limited shareability and distribution of sharing multi-party quantum correlations is used as a fundamental ingredient in many quantum information processing tasks. In quantum cryptography, the amount of information an eavesdropper could possibly obtained about the secret key can be efficiently bounded by using the mutually-exclusive nature of quantum entanglement characterized by monogamy inequalities~\cite{Paw}. Entanglement monogamy also plays an important role in condensed-matter physics such as the $N$-representability problem for fermions~\cite{anti}.

Based on the concept of {\em $q$-expectation value} for real $q>0$, von Neumann entropy can be generalized
into a one-parameter family of entropy functions, namely, {\em Tsallis $q$-entropy}~\cite{tsallis, lv}.
Tsallis $q$-entropy plays an important role in various areas of physics. In nonextensive statistical mechanics, for example, Tsallis-$q$ entropy is theoretically consistent with the minimum cross-entropy principle(or the relative-entropy principle), which excludes the possibility of using the ordinary expectation value from nonextensive statistical mechanics~\cite{LP, Abe}.

Tsallis $q$-entropy is also importantly used in quantum information theory; the concavity of Tsallis-$q$ entropy guarantees the property of {\em entanglement monotone}, a key requirement to construct a faithful entanglement measure~\cite{vidal}. Some conditions on separability criteria of quantum states can also be found by using Tsallis entropy~\cite{ar,tlb,rc}. There are also discussions about characterizing classical statistical correlations inherented in quantum states in terms of $q$-expectation and Tsallis-$q$ entropy~\cite{rr, bpcp}.

In this paper, we show that the polygamy property of multi-party quantum correlations can be characterized in a {\em stronger} form in terms of Tsallis $q$-entropy and $q$-expectation value.
By considering the amount of entanglement that can be distributed in multi-party systems, we establish a class of strong polygamy inequalities of multi-party entanglement in terms of Tsallis $q$-entropy and $q$-expectation.
Our new class of inequalities is in fact tighter(or finer) than the usual polygamy inequalities of multi-party entanglement; the tightness of new inequalities is illustrated by an example. Moreover, our new class of inequalities is concerned with the entanglement distributed between a single party and any possible subsets of the rest parties whereas the usual polygamy inequality only considers the entanglement between one and another. Thus the term {\em strong} is twofold.

We further establish the equivalence between strong polygamy inequalities of quantum entanglement and quantum discord, a different kind of quantum correlation, distributed in multi-party quantum systems using Tsallis $q$-entropy and $q$-expectation. When $q$ tends to 1, Tsallis-$q$ entropy and $q$-expectation are reduced to von Neumann entropy and ordinary expectation, respectively. Thus, our results encapsulate previous results of strong polygamy inequalities of quantum correlations as special cases~\cite{KimEur14}.

This paper is organized as follows. In Section~\ref{Sec: Bipartite Quantum Correlations}, we briefly recall the definitions and properties of $q$-expected correlations in bipartite quantum systems. In Section~\ref{sub: spoly entanglement}, we establish a class of strong polygamy inequalities of entanglement distributed in multi-party quantum systems of arbitrary dimensions in terms of Tsallis-$q$ entropy and $q$-expectation values. In Section~\ref{sub: tightness}, we justify the tightness of our new inequalities by providing an explicitly example. In Section~\ref{spoly discord}, we prove the equivalence between strong polygamy inequalities of quantum entanglement and quantum discord distributed in multi-party quantum systems with respect to Tsallis $q$-entropy and $q$-expectation. We summarize our results in Section~\ref{Sec:Con}.

\section{$q$-expected Correlations in Bipartite Quantum Systems}
\label{Sec: Bipartite Quantum Correlations}
For real parameter $q\geq0$ with $~q \ne 1$, Tsallis $q$-entropy of a quantum state $\rho$ is defined as~\cite{tsallis,lv}
\begin{align}
S_{q}\left(\rho\right)=-\T \rho ^{q} \ln_{q} \rho,
\label{Qtsallis}
\end{align}
where
\begin{eqnarray}
\ln _{q} x &=&  \frac {x^{1-q}-1} {1-q},
\label{qlog}
\end{eqnarray}
is the {\em generalized logarithmic function}(or $q$-logarithm).
Based on the spectrum $\{\lambda_i \}_{i}$ of the quantum state $\rho$~\cite{spec}, Tsallis $q$-entropy in
Eq.~(\ref{Qtsallis}) can also be expressed as
\begin{align}
S_{q}\left(\rho\right)=-\sum_{i}\lambda_{i}^q \ln _{q}\lambda_i,
\label{Ctsallis}
\end{align}
which is the $q$-{\em expectation value} of the $q$-logarithm.
Because $q$-logarithm in Eq.~(\ref{qlog}) converges to the natural logarithm as $q$ tends to $1$,
Tsallis $q$-entropy is a one-parameter generalization of von Neumann entropy based on the concept of $q$-expectation value, in the sense that it converges to the von Neumann entropy
\begin{equation}
\lim_{q\rightarrow 1}S_{q}\left(\rho\right)=-\T\rho \ln \rho.
\end{equation}

Using Tsallis $q$-entropy for $q\geq0$, $q$-{\em expected entanglement}($q$-E)~\cite{Kim19}
of a bipartite pure state $\ket{\psi}_{AB}$ was defined as
\begin{equation}
{E}_{q}\left(\ket{\psi}_{AB} \right)=S_{q}(\rho_A),
\label{qEpure}
\end{equation}
where $\rho_A=\T _{B} \ket{\psi}_{AB}\bra{\psi}$ is the reduced density matrix of $\ket{\psi}_{AB}$ on subsystem $A$.
For a bipartite mixed state $\rho_{AB}$, its $q$-E was defined as the minimum $q$-expectation value
\begin{equation}
E_{q}\left(\rho_{AB} \right)=\min \sum_i p^q_i E_{q}(\ket{\psi_i}_{AB}),
\label{qEmixed}
\end{equation}
whereas $q$-{\em expected entanglement of assistance}($q$-EOA)~\cite{Kim19} was defined as the maximum $q$-expectation value
\begin{equation}
E^a_{q}\left(\rho_{AB} \right)=\max \sum_i p_i^q E_{q}(\ket{\psi_i}_{AB}),
\label{qEOA}
\end{equation}
over all possible pure state decompositions of $\rho_{AB}$,
\begin{equation}
\rho_{AB}=\sum_{i} p_i |\psi_i\rangle_{AB}\langle\psi_i|.
\label{decomp}
\end{equation}

When $q$ tends to 1, Tsallis $q$-entropy converges to von Neumann entropy and the $q$-expectation value becomes ordinary expectation value, therefore, $q$-E and $q$-EOA converge to the entanglement of formation(EOF)~\cite{bdsw} and the entanglement of assistance(EOA)~\cite{cohen}, respectively. Thus, $q$-E and $q$-EOA are one-parameter generalizations of EoF and EOA, respectively, in the full range of non-negative $q$ based on the concept of $q$-expectation value.

Now, we recall more generalized quantum correlations based on $q$-expectation value and Tsallis $q$-entropy.
For a probability ensemble $\mathcal E = \{p_i, \rho_i\}$ of a quantum state $\rho$,
the {\em Tsallis $q$-difference} was defined as~\cite{Kim16T}
\begin{align}
\chi_q\left(\mathcal E\right)=S_q\left(\rho\right)-\sum_{i}p_{i}^q S_q\left(\rho_i\right).
\label{eq: q-diff}
\end{align}
We nota that the concavity of Tsallis $q$-entropy guarantees the non-negativity of Tsallis $q$-difference
for $q\geq 1$, and it converges to the Holevo quantity when $q$ tends to 1.

For a bipartite quantum state $\rho_{AB}$, its {\em one-way classical $q$-correlation}
($q$-CC)~\cite{Kim19} was defined as the maximum Tsallis-$q$ difference
\begin{align}
{\mathcal J}_q^{\leftarrow}(\rho_{AB})&= \max_{\mathcal E} \chi_q\left(\mathcal E\right)
\label{qCC}
\end{align}
over all possible ensemble representations $\mathcal E = \{p_x, \rho_A^x\}$ of the reduced density matrix
$\rho_A$
induced by measurements $\{M^x_B\}$ on subsystem $B$ in such a way that $p_x=\T[(I_A\otimes M_B^x)\rho_{AB}]$
is the probability of the outcome $x$ and
$\rho^x_A=\T_B[(I_A\otimes {M_B^x})\rho_{AB}]/p_x$ is the state of system $A$ when the outcome was $x$.
As a dual quantity to $q$-CC, the {\em one-way unlocalizable $q$-entanglement}($q$-UE)~\cite{Kim16T} was defined by taking the minimum Tsallis-$q$ difference
\begin{align}
{\mathbf u}E_q^{\leftarrow}(\rho_{AB}) &= \min_{\mathcal E} \chi_q\left(\mathcal E\right),
\label{qUE}
\end{align}
over all probability ensembles $\mathcal E $ of $\rho_A$ induced by {\em rank-1 measurements} on subsystem $B$.
We note that $q$-CC and $q$-UE are generalizations of the {\em one-way classical correlation}(CC)~\cite{KW} and
{\em one-way unlocalizable entanglement}(UE)~\cite{BGK}, respectively, due to their convergence when $q$ tends to $1$.

The following proposition shows the trade-off relations between $q$-CC and $q$-E as well as $q$-UE and $q$-EOA distributed in three-party quantum systems.
\begin{Prop}~\cite{Kim19}
For $q \geq 1$ and a three-party pure state $\ket{\psi}_{ABC}$ with its reduced density matrices $\rho_{AB}=\T_C\ket{\psi}_{ABC}\bra{\psi}$,
$\rho_{AC}=\T_B\ket{\psi}_{ABC}\bra{\psi}$ and $\rho_{A}=\T_{BC}\ket{\psi}_{ABC}\bra{\psi}$, we have
\begin{align}
S_q(\rho_A)={\mathcal J}_q^{\leftarrow}(\rho_{AB})+E_q\left(\rho_{AC}\right)
\label{qCCEq}
\end{align}
and
\begin{align}
S_q(\rho_A)={\mathbf u}E_q^{\leftarrow}(\rho_{AB})+E^a_q\left(\rho_{AC}\right).
\label{qUEEqa}
\end{align}
\label{thm: qUEEqa}
\end{Prop}

The concept of $q$-expectation value together with Tsallis-$q$ entropy can also be used to generalize another kind of quantum correlation, {\em quantum discord}~\cite{discord}; based on Tsallis $q$-mutual entropy,
\begin{align}
{\mathcal I}_q\left(\rho_{AB}\right)=S_q\left(\rho_A\right)+S_q\left(\rho_B\right)-
S_q\left(\rho_{AB}\right),
\label{eq: qmutul}
\end{align}
{\em quantum $q$-discord}($q$-D)\cite{qdiscord} of $\rho_{AB}$ was defined as the difference between its Tsallis-$q$ mutual entropy and $q$-CC,
\begin{align}
\delta_q^{\leftarrow}(\rho_{AB})={\mathcal I}_q\left(\rho_{AB}\right)-{\mathcal J}_q^{\leftarrow}(\rho_{AB}).
\label{qdis}
\end{align}
Due to the duality between $q$-CC and $q$-UE, a dual definition to $q$-D was also provided as
\begin{align}
{\mathbf u}\delta_q^{\leftarrow}(\rho_{AB})={\mathcal I}_q\left(\rho_{AB}\right)-{\mathbf u}E_q^{\leftarrow}(\rho_{AB}).
\label{qudis}
\end{align}
Eq.~(\ref{qudis}) is referred to as the {\em one-way unlocalizable quantum $q$-discord}($q$-UD) of $\rho_{AB}$~\cite{Kim19},
which is a generalization of {\em one-way unlocalizable quantum discord}(UD)~\cite{XFL12}.

The following proposition provides a trade-off relation between quantum entanglement($q$-UE) and quantum discord($q$-UD) distributed in three-party quantum systems.
\begin{Prop}~\cite{Kim19}
For $q \geq 1$ and a three-party pure state $\ket{\psi}_{ABC}$ with its reduced density matrices $\rho_{AB}=\T_C\ket{\psi}_{ABC}\bra{\psi}$,
$\rho_{AC}=\T_B\ket{\psi}_{ABC}\bra{\psi}$ and $\rho_{A}=\T_{BC}\ket{\psi}_{ABC}\bra{\psi}$, we have
\begin{align}
S_q(\rho_A)={\mathbf u}\delta_q^{\leftarrow}(\rho_{BA})+{\mathbf u}E_q^{\leftarrow}(\rho_{CA}).
\label{qUEqD}
\end{align}
\label{thm: qUEqD}
\end{Prop}

\section{Strong Polygamy of $q$-expected Quantum Entanglement}
\label{Sec: Strong Polygamy}
\subsection{Strong polygamy inequality of $q$-expected entanglement}
\label{sub: spoly entanglement}

Recently, the polygamous nature of quantum entanglement in multi-party systems was shown to be
characterized in terms of $q$-expectation value~\cite{Kim19_2}; for $q\geq 1$, and
any $(n+1)$-party quantum state $\rho_{AB_1\cdots B_n}$ with bipartite reduced density matrices
$\rho_{AB_i}$ for $i=1,\cdots,n$, we have
\begin{align}
E^a_{q}\left(\rho_{A(B_1\cdots B_n)}\right)
\leq& \sum_{i=1}^{n}E^a_{q}\left(\rho_{AB_i}\right),
\label{Tqpolyn}
\end{align}
conditioned on the superadditivity of Tsallis-$q$ mutual entropy for the classical-classical-quantum(ccq) states,
\begin{align}
{\mathcal I}_q\left(\Omega_{XY:AB_i}\right)\geq {\mathcal I}_q\left(\Omega_{X:AB_i}\right)+{\mathcal I}_q\left(\Omega_{Y:AB_i}\right).
\label{subadd1}
\end{align}
Here, the four-qudit ccq state $\Omega_{XYAB_i}$ is obtained from the two-qudit reduced density matrix $\rho_{AB_i}$ as~\cite{party_qudit}
\begin{align}
\Omega_{XYAB_i}=\frac 1{d^2}\sum_{x,y=0}^{d-1}&\ket{x}_X
\bra{x}\otimes\ket{y}_Y\bra{y}\nonumber\\
&\otimes(I_A\otimes X^x_{B_i}Z^y_{B_i})\rho_{A{B_i}}(I_A\otimes
Z^{-y}_{B_i}X^{-x}_{B_i}),
\label{XYAB}
\end{align}
where
\begin{align}
Z_{B_i}=\sum_{j=0}^{d-1}\omega_d^j\ket{e_j}_B\bra{e_j},~~
X_{B_i}=\sum_{j=0}^{d-1}\ket{e_{j+1}}_B\bra{e_j}
\label{paulis}
\end{align}
are the generalized $d$-dimensional Pauli operators in terms of the eigenvectors
$\{ \ket{e_j }_{B_i}\}$ of $\rho_{B_i}$ and $\omega_d = e^{\frac{2\pi i}{d}}$ is the $d$th-root of unity.

Now, we establish a class of strong polygamy inequalities of multi-party quantum entanglement that deals with the $q$-EOA between between a fixed party and any possible subset of the rest parties.
Let $\B = \{B_1,\cdots,B_n \}$ be the set of subsystems $B_i$'s for $i=1,~\cdots , n$, and
$\X=\{B_{i_1},\cdots,B_{i_k}\}$ be a non-empty proper subset of $\B$ for
$1 \leq k \leq n-1$.
Together with the complement $\X^c=\B-\X$ of $\X$ in $\B$, any $n+1$-party quantum state $\rho_{AB_1\cdots B_n}$
can be considered as a three-party quantum state $\rho_{A\X\X^c}$ with respect to the tripartition $A$-$\X$-$\X^c$.

Conditioned on the the superadditivity of Tsallis-$q$ mutual entropy for ccq states as in (\ref{subadd1}), the polygamy inequality of $q$-EOA leads us to
\begin{align}
E_q^a\left(\rho_{A\B}\right)=&E_q^a\left(\rho_{A(\X\X^c)}\right)\nonumber\\
\leq& E_q^a\left(\rho_{A\X}\right)+E_q^a\left(\rho_{A\X^c}\right),
\label{polyAXXC}
\end{align}
where $E_q^a\left(\rho_{A\X}\right)$ and $E_q^a\left(\rho_{A\X^c}\right)$ are $q$-EOA of reduced density matrices $\rho_{A\X}$ and $\rho_{A\X^c}$, respectively.
By considering all possible non-empty proper subsets $\X$ of $\B$, we have
\begin{align}
E_q^a\left(\rho_{A\B}\right)&\leq \frac{1}{2^n-2}\sum_{\X}\left(E_q^a\left(\rho_{A\X}\right)+E_q^a\left(\rho_{A\X^c}\right)\right),
\label{3polyall}
\end{align}
where the summation is over all possible non-empty proper subsets $\X$'s of $\B$.
Because the set of all non-empty proper subsets of $\B$ is the same with the set of their complements~\cite{subseqcomp},
we have
\begin{align}
\sum_{\X}\left(E_q^a\left(\rho_{A\X}\right)+E_q^a\left(\rho_{A\X^c}\right)\right)=2\sum_{\X}E_q^a\left(\rho_{A\X}\right).
\label{spolyE}
\end{align}

For $1\leq k \leq n-1$ and a non-empty proper subset $\X=\{B_{i_1},\cdots,B_{i_k}\}$ of $\B$ with its complement $\X^c=\{B_{i_{k+1}},\cdots,B_{i_n}\}$,
the polygamy inequality of $q$-EOA in~(\ref{Tqpolyn}) leads us to
\begin{align}
E_q^a\left(\rho_{A\X}\right)\leq\sum_{j=1}^{k}E_q^a\left(\rho_{AB_{i_j}}\right)
\label{upper0}
\end{align}
and
\begin{align}
E_q^a\left(\rho_{A\X^c}\right)\leq \sum_{j=k+1}^{n}E_q^a\left(\rho_{AB_{i_j}}\right).
\label{upper1}
\end{align}
Thus we have
\begin{align}
E_q^a\left(\rho_{A\X}\right)+&E_q^a\left(\rho_{A\X^c}\right)\nonumber\\
&\leq\sum_{j=1}^{k}E_q^a\left(\rho_{AB_{i_j}}\right)+\sum_{j=k+1}^{n}E_q^a\left(\rho_{AB_{i_j}}\right)\nonumber\\
&=\sum_{i=1}^{n}E_q^a\left(\rho_{AB_i}\right).
\label{upper}
\end{align}

Using Inequalities~(\ref{spolyE}), (\ref{upper}) and considering all possible non-empty proper subsets $\X$ of $\B$, we have
\begin{align}
\frac{1}{2^{n-1}-1}\sum_{\X}E_q^a\left(\rho_{A\X}\right)\leq \sum_{i=1}^{n}E_q^a\left(\rho_{AB_i}\right).
\label{upper2}
\end{align}
From inequalities~(\ref{3polyall}) and (\ref{upper2}), we have the following theorem about {\em strong polygamy inequalities}
of $q$-expected entanglement in multi-party quantum systems;
\begin{Thm}
For $q \geq 1$ and any multi-party quantum state $\rho_{AB_1\cdots B_n}=\rho_{A\B}$, we have
\begin{align}
E_q^a\left(\rho_{A\B}\right)\leq\frac{1}{2^{n-1}-1}\sum_{\X}E_q^a\left(\rho_{A\X}\right)
\leq \sum_{i=1}^{n}E_q^a\left(\rho_{AB_i}\right),
\label{spolyE2}
\end{align}
conditioned on the superadditivity of Tsallis-$q$ mutual entropy for the ccq state in (\ref{subadd1}).
Here, the first summation is over all non-empty proper subsets $\X$ of $\B$.
\label{thm: sqpolyE}
\end{Thm}
We note that Inequalities~(\ref{spolyE2}) of Theorem~\ref{thm: sqpolyE} is concerned with the $q$-EOA between the single party $A$ and all possible subsets $\X$'s of $\B$, whereas the $q$-expected polygamy inequality in (\ref{Tqpolyn}) only deals with $q$-EOA between $A$ and each single party ($B_i$'s) in $\B$. We also note that Inequality~(\ref{spolyE2}) is in fact tighter
than the usual $q$-expected polygamy inequality in (\ref{Tqpolyn}).

For $q=1$, the condition in~(\ref{subadd1}) is reduced to the superadditivity of quantum mutual information
\begin{align}
{\mathcal I}\left(\Omega_{XY:AB}\right)\geq {\mathcal I}\left(\Omega_{X:AB}\right)+{\mathcal I}\left(\Omega_{Y:AB}\right),
\label{mutsubadd}
\end{align}
which was shown to be true for any ccq state in general~\cite{Kim16T}.
Thus Inequalities~(\ref{spolyE2}) of Theorem~\ref{thm: sqpolyE} is true for $q=1$ without any condition, and this encapsulates the result of strong polygamy inequality in terms of EOA~\cite{KimEur14}.
\subsection{Tightness}
\label{sub: tightness}

Here, we justify the tightness of our new inequality in~(\ref{spolyE2}) compared with Inequality~(\ref{Tqpolyn}) by providing an example that makes both inequalities in (\ref{spolyE2}) strict; we will consider Inequality~(\ref{spolyE2}) with respect to the {\em Greenberger-Horne-Zeilinger}(GHZ) state~\cite{GHZ}
\begin{equation}
\ket{\rm{GHZ}}_{AB_{1}B_{2}B_{3}}=\frac{1}{\sqrt{2}}\left(\ket{0000}_{AB_{1}B_{2}B_{3}}+
\ket{1111}_{AB_{1}B_{2}B_{3}}\right),
\label{GHZ}
\end{equation}
in four-qubit systems.

Let us first denote $\B=\{B_1, B_2, B_3\}$, so that $\ket{\rm{GHZ}}_{AB_{1}B_{2}B_{3}}=\ket{\rm{GHZ}}_{A\B}$, and consider its reduced density matrix on subsystem $A$
\begin{equation}
\rho_A=\T_{\B}\ket{\rm{GHZ}}_{A\B}\bra{\rm{GHZ}}=\frac{1}{2}\left(\ket{0}_A\bra{0}+\ket{1}_A\bra{1}\right).
\label{GHZrhoA}
\end{equation}
From the definition of $q$-E in (\ref{qEpure}), we can easily obtain the $q$-E of $\ket{\rm{GHZ}}_{A\B}$
with respect to the bipartition between $A$ and $\B$ as
\begin{align}
E_q\left(\ket{\rm{GHZ}}_{A\B}\right)=S_q\left(\rho_A\right)
=\frac{1-2^{1-q}}{q-1},
\label{EoAGHZAB}
\end{align}
which is nonnegative for $q\geq1$. We denote this quantity as
\begin{equation}
\xi_q=\frac{1-2^{1-q}}{q-1}.
\label{EqrhoA}
\end{equation}

For the two-qubit reduced density matrices of $\ket{\rm{GHZ}}_{A\B}$,
we have
\begin{eqnarray}
\rho_{AB_i}&=&\T_{B_jB_k}\ket{\rm{GHZ}}_{A\B}\bra{\rm{GHZ}}\nonumber\\
&=&\frac{1}{2}\left(\ket{00}_{AB_i}\bra{00}+\ket{11}_{AB_i}\bra{11}\right)
\label{GHZrhoABi}
\end{eqnarray}
with distinct $i,~j,~k \in \{1,~2,~3\}$.
Due to the pure-state decomposition of $\rho_{AB_i}$,
\begin{equation}
\rho_{AB_i}=\frac{1}{2}\ket{\phi^{+}}_{AB_i}\bra{\phi^{+}}+\frac{1}{2}\ket{\phi^{-}}_{AB_i}\bra{\phi^{-}}
\label{GHZrhoABiBell}
\end{equation}
in terms of the two-qubit Bell states
\begin{equation}
\ket{\phi^{\pm}}_{AB_i}=\frac{1}{\sqrt{2}}\left(\ket{00}_{AB_i}\pm\ket{11}_{AB_i}\right),
\label{BellABi}
\end{equation}
we can bound the $q$-EOA of $\rho_{AB_i}$ as
\begin{eqnarray}
2^{1-q}\xi_q&=&\left(\frac{1}{2}\right)^qE_q\left(\ket{{\phi^{+}}}_{AB_i}\right)+
\left(\frac{1}{2}\right)^qE_q\left(\ket{{\phi^{-}}}_{AB_i}\right)\nonumber\\
&\leq&E_q^a\left(\rho_{AB_i}\right)\nonumber\\
&\leq& \xi_q,
\label{EoArhoABi}
\end{eqnarray}
where the equality is due to $\T_{B_i}\ket{\phi^{\pm}}_{AB_i}\bra{\phi^{\pm}}=\frac{1}{2}\left(\ket{0}_A\bra{0}+\ket{1}_A\bra{1}\right)$, so that $E_q\left( \ket{\phi^{\pm}}_{AB_i} \right)=\xi_q$, the first inequality is from the definition of $E_q^a\left(\rho_{AB_i}\right)$ that takes the maximum $q$-expectation value, and the last inequality is from the fact that $q$-EOA of any two-qubit state is bounded above by $\xi_q$.
Thus we have
\begin{equation}
E_q^a\left(\rho_{AB_1}\right)=E_q^a\left(\rho_{AB_2}\right)=E_q^a\left(\rho_{AB_3}\right)=\xi
\label{EoArhoABi2}
\end{equation}
for some non-zero $\xi$ such that
\begin{equation}
2^{1-q}\xi_q \leq \xi \leq \xi_q.
\label{xibound}
\end{equation}

Let us now consider the three-qubit reduced density matrices
\begin{eqnarray}
\rho_{AB_iB_j}&=&\T_{B_k}\ket{\rm{GHZ}}_{A\B}\bra{\rm{GHZ}}\nonumber\\
&=&\frac{1}{2}\left(\ket{000}_{AB_iB_j}\bra{000}+\ket{111}_{AB_iB_j}\bra{111}\right),\nonumber\\
\label{GHZrhoAB1B2}
\end{eqnarray}
with distinct $i,~j,~k \in \{1,~2,~3\}$. We note that $\rho_{AB_iB_j}$ in Eq.~(\ref{GHZrhoAB1B2}) can be represented as
\begin{equation}
\rho_{AB_iB_j}=\frac{1}{2}\ket{\phi^{+}}_{AB_iB_j}\bra{\phi^{+}}+\frac{1}{2}\ket{\phi^{-}}_{AB_iB_j}\bra{\phi^{-}},
\label{GHZrhoABiBjBell}
\end{equation}
where
\begin{eqnarray}
\ket{\phi^{\pm}}_{AB_iB_j}
&=&\frac{1}{\sqrt{2}}\left(\ket{000}_{AB_iB_j}\pm\ket{111}_{AB_iB_j}\right)\nonumber\\
\label{GHZ3}
\end{eqnarray}
are the three-qubit GHZ states.

By relabelling the orthonormal basis elements of two-qubit system $B_iB_j$ as
\begin{align}
\ket{00}_{(B_iB_j)}=\ket{\overline{0}}_{(B_iB_j)},~ \ket{11}_{(B_iB_j)}=\ket{\overline{1}}_{(B_iB_j)},
\label{2_3eq}
\end{align}
the three-qubit GHZ states in Eq.~(\ref{GHZ3}) can be rewritten as
\begin{eqnarray}
\ket{\phi^{\pm}}_{A(B_iB_j)}
&=&\frac{1}{\sqrt{2}}\left(\ket{0}_{A}\ket{\overline{0}}_{(B_iB_j)}\pm\ket{1}_{A}\ket{\overline{1}}_{(B_iB_j)}\right),\nonumber\\
\label{BellABi2}
\end{eqnarray}
which are equivalent to the two-qubit bell states in Eq.~(\ref{BellABi}) with respect to the bipartition between $A$ and $B_iB_j$.

Eqs.~(\ref{GHZrhoABiBjBell}) and (\ref{BellABi2}) impliy that the three-qubit reduced density matrices in (\ref{GHZrhoABiBjBell}) are equivalent to the two-qubit states in~(\ref{GHZrhoABiBell}) when we consider the bipartition between $A$ and $B_iB_j$.
Thus the same argument in Inequalities~(\ref{EoArhoABi}), (\ref{xibound}) and Eq.~(\ref{EoArhoABi2}) leads us to
\begin{eqnarray}
E_q^a\left(\rho_{A(B_1B_2)}\right)&=&E_q^a\left(\rho_{A(B_1B_3)}\right)\nonumber\\
&=&E_q^a\left(\rho_{A(B_2B_3)}\right)=\xi.
\label{EoArhoABiBj}
\end{eqnarray}

For the four-qubit GHZ state $\ket{\rm{GHZ}}_{A\B}$, Eq.~(\ref{EoArhoABi2}) and Eq.~(\ref{EoArhoABiBj}) lead the intermediate term of Inequalities~(\ref{spolyE2}) as
\begin{equation}
\frac{1}{2^{n-1}-1}\sum_{\X}E_q^a\left(\rho_{A\X}\right)
=\frac{1}{2^{3-1}-1}6\xi
=2\xi
\label{interme}
\end{equation}
where the summation is over all possible non-empty proper subsets $\X$ of $\B=\{B_1, B_2, B_3\}$.
We also note that the first and last terms of Inequalities~(\ref{spolyE2}) for $\ket{\rm{GHZ}}_{A\B}$ are
\begin{eqnarray}
E_q^a\left(\ket{\rm{GHZ}}_{A\B}\right)=\xi_q
\label{first}
\end{eqnarray}
and
\begin{eqnarray}
\sum_{i=1}^{3}E_q^a\left(\rho_{AB_i}\right)=3\xi,
\label{last}
\end{eqnarray}
respectively.

Now we have
\begin{eqnarray}
\frac{1}{2^{n-1}-1}\sum_{\X}E_q^a\left(\rho_{A\X}\right)
&\geq&2^{2-q}\xi_q\nonumber\\
&>&\xi_q\nonumber\\
&=&E_q^a\left(\ket{\rm{GHZ}}_{A\B}\right)
\label{firint}
\end{eqnarray}
where the first inequality is due to Eq.~(\ref{interme}) as well as Inequality~(\ref{xibound}), and the second inequality is guaranteed to be strict for $1\leq q<2$.
Moreover, Eqs.~(\ref{interme}) and (\ref{last}) lead us to the strict inequality
\begin{eqnarray}
\sum_{i=1}^{3}E_q^a\left(\rho_{AB_i}\right)>\frac{1}{2^{n-1}-1}\sum_{\X}E_q^a\left(\rho_{A\X}\right)
\label{intlas}
\end{eqnarray}
for any $q\geq 1$.
Thus Inequality~(\ref{firint}) and Inequality~(\ref{intlas}) imply that every inequality in~(\ref{spolyE2}) is strict for the four-qubit GHZ state in Eq.~(\ref{GHZ}) in the range of  $1\leq q <2$.

\section{Equivalence in Strong Polygamy of multi-party $q$-expected quantum correlations}
\label{spoly discord}

Recently, the equivalence of polygamy inequalities based on different kind of $q$-expected quantum correlations was established in three-party quantum systems;
\begin{Prop}~\cite{Kim19}
For $q\geq 1$, and any three-party pure state $\ket{\psi}_{ABC}$,
the polygamy inequality of $q$-EoA is equivalent to the polygamy inequality of $q$-UD, that is,
\begin{align}
E_q\left(\ket{\psi}_{A(BC)}\right)
\leq& E_q^a(\rho_{AB})+E_q^a(\rho_{AC})
\label{EoApoly3}
\end{align}
if and only if
\begin{align}
{\mathbf u}\delta_q^{\leftarrow}\left(\ket{\psi}_{A(BC)}\right)
\leq& {\mathbf u}\delta_q^{\leftarrow}(\rho_{AB})+{\mathbf u}\delta_q^{\leftarrow}(\rho_{AC}),
\label{Dqpoly3}
\end{align}
where ${\mathbf u}\delta_q^{\leftarrow}\left(\ket{\psi}_{A(BC)}\right)$ is the $q$-UD of the pure state $\ket{\psi}_{ABC}$ with respect to the bipartition between $A$ and $BC$.
\label{prop: Dqpoly3}
\end{Prop}
Here, we generalize Proposition~\ref{prop: Dqpoly3} in a stronger form that is valid in any multi-party quantum systems.

From the definition of $q$-UD in Eq.~(\ref{qudis}) and the Tsallis $q$-mutual entropy in Eq.~(\ref{eq: qmutul}) as well as the trade-off relation between $q$-EU and $q$-EOA in Eq.~(\ref{qUEEqa}) of Proposition~\ref{thm: qUEEqa}, we have
\begin{align}
E_q^a\left(\rho_{A\X}\right)={\mathbf u}{\delta}_q^{\leftarrow}\left(\rho_{A\X^c}\right)+S_q\left(\rho_{A|\X^c}\right).
\label{EDEoA2}
\end{align}
for any multi-party pure state $\ket{\psi}_{A\B}=\ket{\psi}_{AB_1\cdots B_n}$ and a non-empty proper subset $\X$ of $\B$,
where $S_q\left(\rho_{A|{\X}^c}\right)=S\left(\rho_{A{\X}^c}\right)-S\left(\rho_{{\X}^c}\right)$ is the Tsallis $q$-conditional entropy of $\rho_{A{\X}^c}$.
Thus we have
\begin{align}
\sum_{\X}E_q^a\left(\rho_{A\X}\right)
=&\sum_{\X}{\mathbf u}\delta_q^{\leftarrow}\left(\rho_{A\X^c}\right)+\sum_{\X}S_q\left(\rho_{A|\X^c}\right)\nonumber\\
=&\sum_{\X}{\mathbf u}\delta_q^{\leftarrow}\left(\rho_{A\X}\right)+\sum_{\X}S_q\left(\rho_{A|\X}\right),
\label{EDEoAsum}
\end{align}
where the summations are over all possible non-empty proper subsets $\X$ of $\B$, and the last equality is
due to~\cite{subseqcomp}.

Because of the complementary property of Tsallis $q$-conditional entropy~\cite{compqcond}, we have
\begin{align}
S_q\left(\rho_{A|\X}\right)+S_q\left(\rho_{A|\X^c}\right)=0
\label{compcondent}
\end{align}
for any multi-party pure state $\ket{\psi}_{A\B}$ and any non-empty proper subset $\X$ of $\B$.
Thus we have
\begin{align}
\sum_{\X}S_q\left(\rho_{A|\X}\right)=0,
\label{compcondent2}
\end{align}
with the summation over all non-empty proper subsets $\X$ of $\B$, and this leads Eqs.~(\ref{EDEoAsum}) to
\begin{align}
\sum_{\X}E_q^a\left(\rho_{A\X}\right)=\sum_{\X}{\mathbf u}\delta_q^{\leftarrow}\left(\rho_{A\X}\right),
\label{sumEDEU}
\end{align}
for any multi-party pure state $\ket{\psi}_{A\B}$.

Now let us consider $\ket{\psi}_{A\B}$ together with its environmental system $\ket{\phi}_C$.
Due the the universality of Eq.~(\ref{qUEqD}) in Proposition~\ref{thm: qUEqD}  considered on $\ket{\psi}_{A\B}\otimes \ket{\phi}_C$, we have
\begin{align}
S_q(\rho_{\B})
=&{\mathbf u}\delta_q^{\leftarrow}(\ket{\psi}_{A\B})+{\mathbf u}E_q^{\leftarrow}\left(\ket{\phi}_C\bra{\phi}\otimes \rho_{\B}\right),
\label{qUEqD2}
\end{align}
where $\rho_{\B}=\T_A \ket{\psi}_{A\B}\bra{\psi}$.
As the subsystems ${\B}C$ of $\ket{\psi}_{A\B}\otimes \ket{\phi}_C$ is in a product state, any measurement on subsystem $\B$ leaves the subsystem $C$ intact, that is, $\ket{\phi}_C\bra{\phi}$. Thus, the definition of $q$-UE in~(\ref{qUE}) leads us to
\begin{align}
{\mathbf u}E_q^{\leftarrow}\left(\ket{\phi}_C\bra{\phi}\otimes \rho_{\B}\right)
=&\min \left[S_q (\ket{\phi}_C)-\sum_x p_x^q S_q(\ket{\phi}_C)\right]\nonumber\\
=&0.
\label{qud0}
\end{align}
From Eqs.~(\ref{qUEqD2}) and (\ref{qud0}), we have
\begin{align}
{\mathbf u}\delta_q^{\leftarrow}(\ket{\psi}_{A\B})=S_q(\rho_{\B}),
\label{qUEqD3}
\end{align}
therefore
\begin{align}
{\mathbf u}\delta_q^{\leftarrow}(\ket{\psi}_{A\B})=S_q(\rho_{\B})=S_q(\rho_A)=E_q\left(\ket{\psi}_{A{\B}}\right),
\label{qUEqD3sam}
\end{align}
for any multi-party pure state $\ket{\psi}_{A\B}$ with respect to the bipartition between $A$ and $\B$..

From Eqs.~(\ref{sumEDEU}) and (\ref{qUEqD3sam}) together with
Inequality~(\ref{spolyE2}), we have the following theorem about the equivalence of strong polygamy
inequalities in terms of $q$-EOA and $q$-UD in multi-party quantum systems.

\begin{Thm}
For $q \geq 1$ and any multi-party quantum state $\ket{\psi}_{AB_1\cdots B_n}=\ket{\psi}_{A\B}$, the strong polygamy inequality
of $q$-EOA is equivalent to that of $q$-UD, that is,
\begin{align}
E_q\left(\ket{\psi}_{A\B}\right)\leq\frac{1}{2^{n-1}-1}\sum_{\X}E_q^a\left(\rho_{A\X}\right),
\label{sqpolyE2}
\end{align}
if and only if
\begin{align}
{\mathbf u}\delta_q^{\leftarrow}\left(\ket{\psi}_{A\B}\right)\leq\frac{1}{2^{n-1}-1}\sum_{\X}{\mathbf u}\delta_q^{\leftarrow}\left(\rho_{A\X}\right),
\label{sqpolyD}
\end{align}
where the summations are over all non-empty proper subsets $\X$ of $\B$.
\label{the: sqpeq}
\end{Thm}

Theorem~\ref{the: sqpeq} is a generalization of Proposition~\ref{prop: Dqpoly3} into multi-party quantum systems
in a stronger form that deals with quantum correlations between the single party $A$ and all possible non-empty proper subsets $\X$'s of $\B$. We further note that, due to Theorem~\ref{thm: sqpolyE}, Inequality~(\ref{sqpolyE2}) itself is guaranteed on the condition of the superadditivity of Tsallis-$q$ mutual entropy for the ccq states in (\ref{subadd1}), therefore Inequality~(\ref{sqpolyD}) itself is also guaranteed on the same condition. However, Theorem~\ref{the: sqpeq} says that the equivalence of the strong polygamy inequalities in terms of $q$-EOA and $q$-UD is always guaranteed with no condition.

\section{Summary}
\label{Sec:Con}
Using Tsallis $q$-entropy and $q$-expectation value, we have shown that the polygamous nature of multi-party quantum correlations can be characterized in a stronger form; we have established a class of strong polygamy inequalities of multi-party entanglement in terms of $q$-EOA. Our new class of inequalities was shown to be tighter than the usual polygamy inequalities of multi-party entanglement, which was illustrated by an explicit example. We also note that our new class of inequalities is concerned with the entanglement distributed between a single party and any possible subsets of the rest parties whereas the usual polygamy inequality only considers the entanglement between one and another. Thus the term `strong' can be doubly implied here.

We have also established the equivalence between strong polygamy inequalities of different kind of quantum correlations, quantum entanglement and quantum discord, in terms of $q$-EOA and $q$-UD. When $q$ tends to 1, Tsallis-$q$ entropy and $q$-expectation are reduced to von Neumann entropy and ordinary expectation, respectively. Thus, our results encapsulate previous results of strong polygamy inequalities of quantum correlations as special cases.

As the limited shareability of quantum correlations in multi-party quantum systems is the key ingredient in many quantum information tasks, it is important and even necessary to have proper mathematical characterizations of these trade-off relations such as monogamy and polygamy inequalities of quantum correlations. We also note that study of quantum mechanical properties in higher-dimensional systems is sometimes preferred in many quantum cryptographical protocols, because it provides us with higher coding density and stronger security.

Our results here strengthen the characterization of the polygamous nature in distributing quantum correlations in multi-party high-dimensional quantum systems, which can be useful methods for the foundation of many secure quantum information and communication protocols. Noting the importance of the study on multi-party quantum correlations, our results here can also provide a rich reference for future work to understand the nature of multi-party quantum correlations.

\section*{Acknowledgments}
This work was supported by Basic Science Research Program(NRF-2020R1F1A1A010501270) and Quantum Computing Technology Development Program(NRF-2020M3E4A1080088) through the National Research Foundation of Korea(NRF) grant funded by the Korea government(Ministry of Science and ICT).



\begin{thebibliography}{1}
\bibitem{party_qudit}
Without loss of generality, we may assume the ($n+1$)-party state $\rho_{AB_1\cdots B_n}$ is an ($n+1$)-qu$d$it state
by taking $d$ as the largest dimension of the subsystems.
\end{thebibliography}
\end{document}